\documentclass[twocolumn,english,aps,pra,superscriptaddress,showpacs,tightenlines]{revtex4-1}
\usepackage{amsmath}
\usepackage{graphicx}
\usepackage{amssymb}

\begin{document}

%\begin{CJK*}{GBK}{song}

\title{Concurrence of Two Identical Atoms in a Rectangular Waveguide: Linear Approximation with Single Excitation}

\author{Lijuan \surname{Hu}}
\affiliation{Key Laboratory of Low-Dimensional Quantum Structures and Quantum Control of
Ministry of Education, Department of Physics and Synergetic Innovation
Center of Quantum Effects and Applications, Hunan Normal University,
Changsha 410081, China}
\author{Guiyuan \surname{Lu}}
\affiliation{Key Laboratory of Low-Dimensional Quantum Structures and Quantum Control of
Ministry of Education, Department of Physics and Synergetic Innovation
Center of Quantum Effects and Applications, Hunan Normal University,
Changsha 410081, China}
\author{Jing \surname{Lu}}
\affiliation{Key Laboratory of Low-Dimensional Quantum Structures and Quantum Control of
Ministry of Education, Department of Physics and Synergetic Innovation
Center of Quantum Effects and Applications, Hunan Normal University,
Changsha 410081, China}
\author{Lan \surname{Zhou}}
\thanks{Corresponding author}
\email{zhoulan@hunnu.edu.cn}
\affiliation{Key Laboratory of Low-Dimensional Quantum Structures and Quantum Control of
Ministry of Education, Department of Physics and Synergetic Innovation
Center of Quantum Effects and Applications, Hunan Normal University,
Changsha 410081, China}

\begin{abstract}
We study two two-level systems (TLSs) interacting with a reservoir of guided modes confined
in a rectangular waveguide. For the energy separation of the identical TLSs far away from the
cutoff frequencies of transverse modes, the delay-differential equations are obtained with single
excitation initial in the TLSs. The effects of the inter-TLS distance on the time evolution
of the concurrence of the TLSs are examined.
\end{abstract}

\pacs{03.65.Yz, 03.65.-w}
%03.65.Yz 	Decoherence; open systems;
%03.65.-w 	Quantum mechanics

\maketitle

%\end{CJK*}\narrowtext

\section{Introduction}

Quantum entanglement is a nonlocal correlation of multipartite quantum
systems, which distinguishes the quantum world from the classical world. Due
to its important role in quantum computation and communication, it is a
physical resource which quantum technologies are based on. However, the
inevitable interaction of quantum systems with their surrounding
environments induces decoherence of quantum systems, which degrades the
entanglement of quantum systems. Understanding the dynamics of the
entanglement is desirable to be able to manipulate entanglement states in a
practical way as well as the question of emergent classicality from quantum
theory. Entanglement dynamics is studied under local decoherence (two
particles in an entangled state are coupled to its own environment
individually), a peculiar dynamical feature of entangled state is that
complete disentanglement is achieved in finite time although complete
decoherence takes an infinite time, which is termed ``entanglement sudden
death''~\cite{ESD1,ESD2}. The assumption of local decoherence requires that
two two-level systems (TLSs), e.g. atoms, are sufficiently separated. It is
well known that the radiation field emitted by an atom may influence the
dynamics of its closely spaced atoms~\cite%
{Dicke,Lehmberg70,Feng41,Ordon70,BermanPRA76}. The entanglement can be
generated in a two-atom system after a finite time by their cooperative
spontaneous emission, or the destroyed entanglement may reappear suddenly
after its death, which is known as sudden birth of entanglement~\cite{ESB}.

In quantum network, stationary qubits generate, store, and process quantum
information at quantum nodes, and flying qubit transmit quantum information
between the nodes through quantum channels. A distributed quantum network
requires coherently transferring quantum information among stationary
qubits, flying qubits, and between stationary qubits and flying qubits. With
the development of techniques in quantum information, an alternative
waveguide-based quantum electrodynamics (QED) system has emerged as a
promising candidate for achieving quantum network~\cite{Tien,HamRMP82,Caruso}%
. In this system, atoms are located at quantum nodes and photons propagating
along the network are confined in a waveguide. Inside a one-dimensional (1D)
waveguide, the electromagnetic field is confined spatially in two dimensions
and propagates along the remaining one, which is called guided modes. The
spectrum of the guided modes is continuous. The coupling of the
electromagnetic field to a TLS can be increased by reducing the transverse
size of the guided modes. Therefore the study of entanglement dynamics in
systems embedded in waveguides is of importance. A waveguide with a cross
section has many guided modes~\cite{TETMmode}, e.g. transverse-magnetic (TM)
modes or transverse-electric (TE) ones. However, most work only consider one
guided mode of the waveguide~\cite%
{Fans,ZLPRL08,ZLQrouter,Zheng,LawPRA78,TShiSun,PRA14Red,Ordonez}. In this
paper, we consider the dynamic behavior of bipartite entanglement involving
two identical TLSs which is implanted into the 1D rectangular hollow
metallic waveguide. Since local addressing is difficult, we assume that
there is no direct interaction between the TLSs, the TLSs and the field
share initially a single excitation. By considering the energy separation of
the TLSs is far away from the cutoff frequencies of the transverse modes,
the delay differential equations are obtained for two TLSs' amplitudes with
the field initially in vacuum, where multiple guided modes are included. The
spatial separation of the two TLSs introduces the position-dependent phase
factor and the time delay (finite time required for light to travel from one
TLS to the other) in each transverse mode. The phase factors and the time
delays are different in different transverse modes. The effect of the phase
factors and the time delays on the entanglement dynamics of the TLSs are
studied in details by considering the TLSs interacting with single
transverse mode and double transverse modes.

This paper is organized as follows. In Sec.~\ref{Sec:2}, we introduce the
model and establish the notation. In Sec.~\ref{Sec:3}, we derive the
relevant equations describing the dynamics of the system for the TLSs being
initially excited and the waveguide mode in the vacuum state, and
investigate the effect of spatial separation on the dynamics of entanglement
between two identical TLSs, which is characterized by concurrence. We make a
conclusion in Sec.~\ref{Sec:4}.

%%%%%%%%%%%%%%%%%%%%%%%%%%%%%%%%%%%%%%%%%%%%%%%%%%%%%%%%%%%%

\section{\label{Sec:2}Two TLSs in a rectangular waveguide}

%%%%%%%%%%%%%%%%%%%%%%%%%%%%%%%%%%%%%%%%%%%%%%%%%%%%%%%%%%%%

\begin{figure}[tbp]
\includegraphics[clip=true,height=6cm,width=8cm]{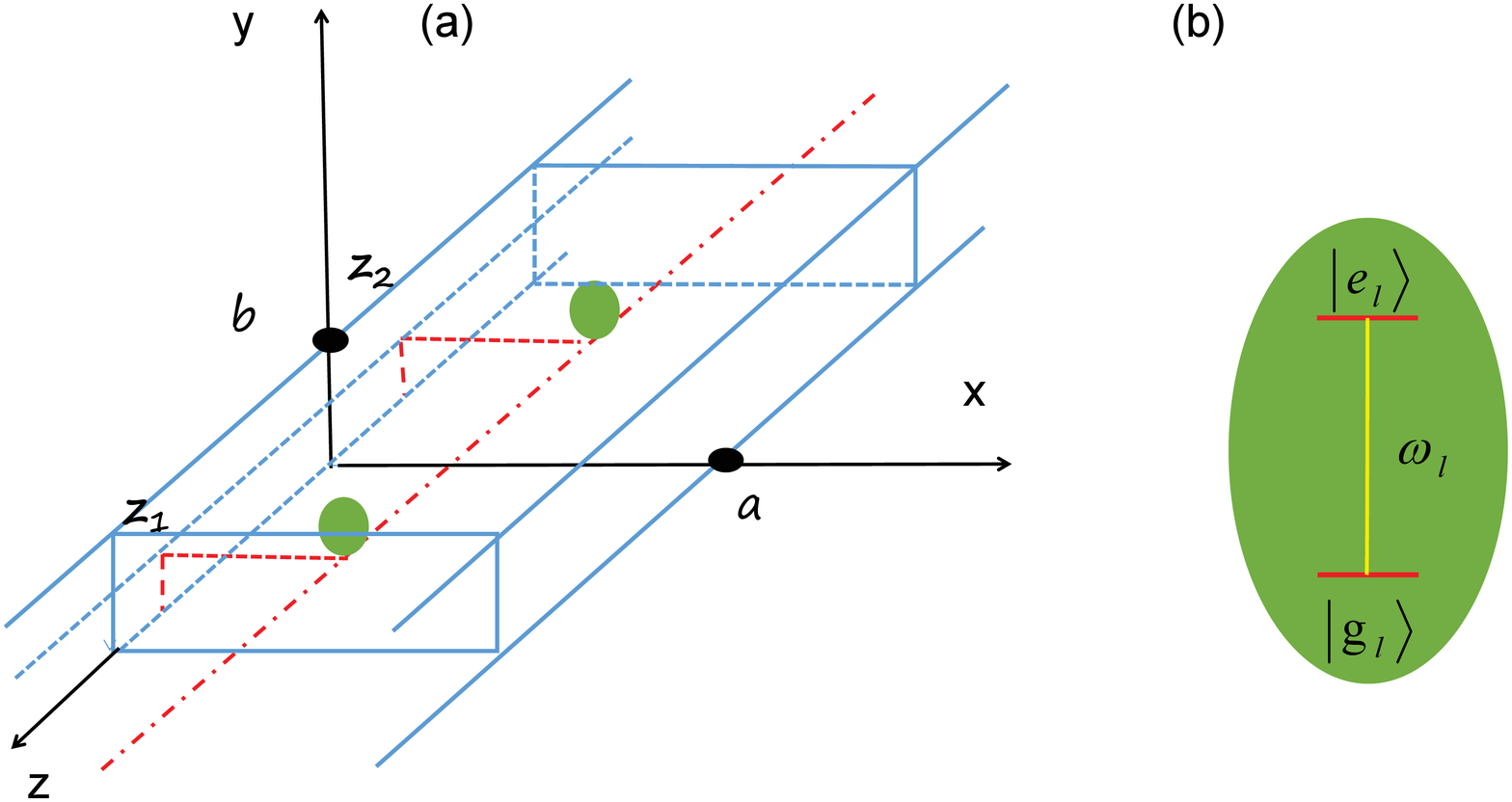}
\caption{(color online) Schematic illustration for an infinite waveguide of
rectangular cross section $A=ab$ (a) coupling to two TLSs (b) located at $%
\vec{r}_{1}=(a/2,b/2,z)$ and $\vec{r}_{2}=(a/2,b/2,z_2)$.}
\label{Fig1.eps}
\end{figure}

We consider a rectangular hollow metallic waveguide with the area $A=ab$ ($%
a=2b$) of its cross section, as shown in Fig.~\ref{Fig1.eps}. The axes of
the waveguide parallel to the $z$ axis, and the waveguide is infinite along
the $z$ axis. Since the translation invariance is maintained along the $z$
axis, all the components of the electromagnetic field describing the guided
mode depend on the coordinate z as $e^{ikz}$. The guided mode can be
characterized by three wave numbers $\{k_{x},k_{y},k_{z}\}$. The spatial
confinement of the electromagnetic field along the $xy$ plane makes the
appearance of the two non-negative integers $m$ and $n$, which related to
wave numbers along the $x$ and $y$ directions by $k_{x}=m\pi /a$ and $%
k_{y}=n\pi /b$. In this waveguide, there are two types of guiding modes~\cite{TETMmode,HuangPRA,Shahmoon,liqongPRA,zhouCTP69}:the transverse-magnetic
modes $TM_{mn}$ ($H_{z}=0$) and the transverse-electric modes $TE_{mn}$ ($%
E_{z}=0$). Two idential TLSs, named TLS $1$ and TLS $2$, are separately
located inside the waveguide at positions $\vec{r}_{1}=(a/2,b/2,z_{1})$ and $%
\vec{r}_{2}=(a/2,b/2,z_{2})$, the distance between the TLSs is denoted by $%
d=z_{2}-z_{1}$. The free Hamiltonian of the TLSs read
\begin{equation}
H_{a}=\sum\limits_{l=1}^{2}\hbar \omega _{A}\sigma _{l}^{+}\sigma _{l}^{-}
\label{2-A1}
\end{equation}%
where $\omega _{A}$ are the energy difference between the excited state $%
|e\rangle $ and the ground state $|g\rangle $, and $\sigma _{l}^{+}\equiv
\left\vert e_{l}\right\rangle \left\langle g_{l}\right\vert $ ($\sigma
_{l}^{-}\equiv \left\vert g_{l}\right\rangle \left\langle e_{l}\right\vert $
) is the rising (lowing) atomic operator of the $l$-$th$ TLS. We assume the
dipoles of TLSs are along the $z$ axis. In this case, only the $TM_{mn}$
guided modes are interacted with the TLSs. The free Hamiltonian of the field
reads
\begin{equation}
H_{f}=\sum_{j}\int dk\hbar \omega _{jk}\hat{a}_{jk}^{\dagger }\hat{a}_{jk}
\label{2-A2}
\end{equation}%
where $\hat{a}_{jk}^{\dagger }$ ($\hat{a}_{jk}$) is the creation
(annihilation) operator of the $TM_{mn}$ modes. Here,we have replaced $(m,n)$
with the sequence number\ $j$, i.e., $j=1,2,3...$ denoting $%
TM_{11},TM_{31},TM_{51}\cdots$, respectively. For each guided mode, the
dispersion relation is given by $\omega _{jk}=\sqrt{\Omega
_{j}^{2}+c^{2}k^{2}} $, where $\Omega _{mn}=c\sqrt{(m\pi /a)^{2}+(n\pi
/b)^{2}}$ is the cutoff frequency. No electromagnetic field can be guided if
their frequency is smaller than the cutoff frequency $\Omega _{1}$. The
interaction between the TLSs and the the electromagnetic field is written as
\begin{equation}
H_{int}=\sum_{l=1}^{2}\sum_{j}\int dk\hbar \frac{g_{jl}}{\sqrt{\omega _{jk}}}%
e^{ikz_{l}}S_{l}^{-}\hat{a}_{k}^{\dagger }+h.c.  \label{2-A3}
\end{equation}%
in the electric dipole and rotating wave approximations, where $%
g_{jl}=\Omega _{j}\mu _{l}/\sqrt{A\pi \epsilon _{0}}$ and $\mu _{l}$ the
magnitude of the dipole of the $l$-th TLS. We assume that $\mu _{1}=\mu
_{2}=\mu $ is real. Then the parameter $g_{jl}$ becomes
\begin{equation}
g_{j}=\frac{\Omega _{j}\mu \sin \left( \frac{m\pi }{2}\right) \sin \left(
\frac{n\pi }{2}\right) }{\sqrt{\hbar A\pi \epsilon _{0}}},  \label{2-A4}
\end{equation}%
where $\epsilon _{0}$ is the permittivity of free space. The TLS's position
is presented in the exponential function in Eq.(\ref{2-A3}).
The total system, the two TLSs and the photons in quantum electromagnetic
field, is described by the Hamiltonian
\begin{equation}
H=H_{f}+H_{a}+H_{int}  \label{2-A5}
\end{equation}%
The total system is a closed system. However, each subsystem is an open
system. When we are only interested in the dynamics of TLSs, the quantum
electromagnetic field can be regarded as an environment.

%%%%%%%%%%%%%%%%%%%%%%%%%%%%%%%%%%%%%%%%%%%%%%%%%%%%%%%%%%%%

\section{\label{Sec:3} Entanglement Dynamics}

Any state of the two TLSs are linear superposition of the basis of the
separable product states $\left\vert 1\right\rangle =\left\vert
g_{1}g_{2}\right\rangle $, $\left\vert 2\right\rangle =\left\vert
e_{1}g_{2}\right\rangle $, $\left\vert 3\right\rangle =\left\vert
g_{1}e_{2}\right\rangle $, and $\left\vert 4\right\rangle =\left\vert
e_{1}e_{2}\right\rangle $. Since the number of quanta is conserved in this
system, the wavefunction of the total system can be written as:%
\begin{equation}
\left\vert \psi (t)\right\rangle =b_{1}\left\vert 20\right\rangle
+b_{2}\left\vert 30\right\rangle +\sum_{j}\int dkb_{jk}a_{jk}^{\dagger
}\left\vert 10\right\rangle  \label{2-A6}
\end{equation}%
in single excitation subspace, where $\left\vert 0\right\rangle $ is the
vacuum state of the quantum field. The first term in Eq.(\ref{2-A6})presents
TLS $1$ in the excited state with no excitations in the field, $b_{1}\left(
t\right) $ is the corresponding amplitude, the second term in Eq.(\ref{2-A6}%
) presents TLS $2$ in the excited state with photons in the vacuum, whereas
the third term in Eq.(\ref{2-A6}) describes all TLSs in the ground state
with a photon emitted at a mode $k$ of the TM$_{j}$ guided mode, $%
b_{jk}\left( t\right) $ is the corresponding amplitude. The initial state of
the system is denoted by the amplitudes $b_{1}\left( 0\right) ,b_{2}\left(
0\right) $, $b_{jk}\left( 0\right) =0$. The Schr\"{o}dinger equation results
in the following coupled equation of the amplitudes
\begin{subequations}
\label{2-A7}
\begin{eqnarray}
\dot{b}_{1} &=&-i\omega _{A}b_{1}-\sum_{j}\int dk\frac{b_{jk}g_{j}}{\sqrt{%
\omega _{jk}}}e^{-ikz_{1}} \\
\dot{b}_{2} &=&-i\omega _{A}b_{2}-\sum_{j}\int dk\frac{b_{jk}g_{j}}{\sqrt{%
\omega _{jk}}}e^{-ikz_{2}} \\
\dot{b}_{jk} &=&-i\omega _{jk}b_{jk}+\frac{g_{j}e^{ikz_{1}}}{\sqrt{\omega
_{jk}}}\left( b_{1}+b_{2}e^{ikd}\right)
\end{eqnarray}%
We introduce three new variables to remove the high-frequency effect
\end{subequations}
\begin{subequations}
\label{2-A8}
\begin{eqnarray}
b_{1}(t) &=&B_{1}(t)e^{-i\omega _{A}t}, \\
b_{2}(t) &=&B_{2}(t)e^{-i\omega _{A}t}, \\
b_{jk}(t) &=&B_{jk}\left( t\right) e^{-i\omega _{jk}t},
\end{eqnarray}%
then, formally integrate equation of $B_{jk}\left( t\right) $, which is
later inserted into the equations for $B_{1}\left( t\right) $ and $%
B_{2}\left( t\right) $. The probability amplitude for one TLS being excited
is determined by two coupled integro-differential equations. Assuming that
the frequency $\omega _{A}$ is far away from the cutoff frequencies $\Omega
_{j}$, we can expand $\omega _{jk}$ around $\omega _{A}$ up to the linear
term
\end{subequations}
\begin{equation}
\omega _{jk}=\omega _{A}+v_{j}\left( k-k_{j0}\right) ,  \label{2-A10}
\end{equation}%
where the wavelength of the emitted radiation $k_{j0}=\sqrt{\omega
_{A}^{2}-\Omega _{j}^{2}}/c$ is determined by $\omega _{jk_{0}}=\omega _{A}$%
, and the group velocity
\begin{equation}
v_{j}\equiv \frac{d\omega _{jk}}{dk}|_{k=k_{j0}}=\frac{c\sqrt{\omega
_{A}^{2}-\Omega _{j}^{2}}}{\omega _{A}}  \label{2-A11}
\end{equation}%
is different for different TM$_{j}$ guided modes. Integrating over all wave
vectors $k$ gives rise to a linear combination of $\delta \left( t-\tau
-\tau _{j}\right) $ and $\delta \left( t-\tau_j \right) $, where $\tau
_{j}=d/v_{j}$ is the time delay taking by a photon traveling from one TLS to
the other TLS in the given transverse mode $j$. The differential equations
governing the dynamics of two TLSs read
\begin{subequations}
\label{2-A12}
\begin{eqnarray}
\left( \partial _{t}+\gamma \right) B_{1}(t) &=&-\sum_{j}\gamma
_{j}e^{i\varphi _{j}}B_{2}\left( t-\tau_{j}\right) \Theta \left(
t-\tau_{j}\right) \\
\left( \partial _{t}+\gamma \right) B_{2}(t) &=&-\sum_{j}\gamma
_{j}e^{i\varphi _{j}}B_{1}\left( t-\tau_{j}\right) \Theta \left(
t-\tau_{j}\right)
\end{eqnarray}%
where we have defined the phase $\varphi _{j}=k_{j0}d$ due to the distance
between the TLSs, the decay rate $\gamma _{j}=\pi \left\vert
g_{j}\right\vert ^{2}/(v_{j}\omega _{A})$ caused by the interaction between
the TLSs and the vacuum field in a give transverse mode $j$, $\Theta(x)$ is
the Heaviside unit step function, i.e., $\Theta(x)=1$ for $x>0$, and $%
\Theta(x)=0$ for $x<0$. The decay to all $TM_{j}$ modes is denoted by $%
\gamma =\sum_{j}\gamma _{j}$, the retard effect~\cite%
{Milonni74,cookPRA53,DungPRA59,DornPRA66,RistPRA78,GulfPRA12,JingPLA377}
has been implied by the symbol $\tau _{j}$. At times less than minimum $\tau
_{j}$, two TLSs decay as if they are isolated in rectangular waveguide.
After the time $\min\tau _{j}$ the TLS recognizes the other TLS due to its
absorption of photons. As time goes on, reemissions and reabsorptions of
photons by two TLSs might produce interference, which leads to the change of
atomic upper state population. It is convenient to write Eq.(~\ref{2-A12})
in the Dicke symmetric state $|s\rangle =(|2\rangle +|3\rangle )/\sqrt{2}$
and antisymmetric state $|a\rangle =(|2\rangle -|3\rangle )/\sqrt{2}$
\end{subequations}
\begin{subequations}
\label{2-A13}
\begin{eqnarray}
\left( \partial _{t}+\gamma \right) C_{s}(t) &=&-\sum_{j}\gamma
_{j}e^{i\varphi _{j}}C_{s}\left( t-\tau_{j}\right) \Theta \left(
t-\tau_{j}\right) \\
\left( \partial _{t}+\gamma \right) C_{a}(t) &=&\sum_{j}\gamma
_{j}e^{i\varphi _{j}}C_{a}\left( t-\tau_{j}\right) \Theta \left(
t-\tau_{j}\right)
\end{eqnarray}%
which allow either TLS 1 or TLS 2 to be excited with equal probability. They
are degenerate eigenstates of Hamiltonian $H_a$. The equations for the
amplitudes of the Dicke states are not coupled.
%%%%%%%%%%%%%%%%%%%%%%%%%%%%%%%%%%%%%%%%%%%%%%%%%%%%%%%%%%%%
\begin{figure}[tbp]
\includegraphics[clip=true,height=17cm,width=8cm]{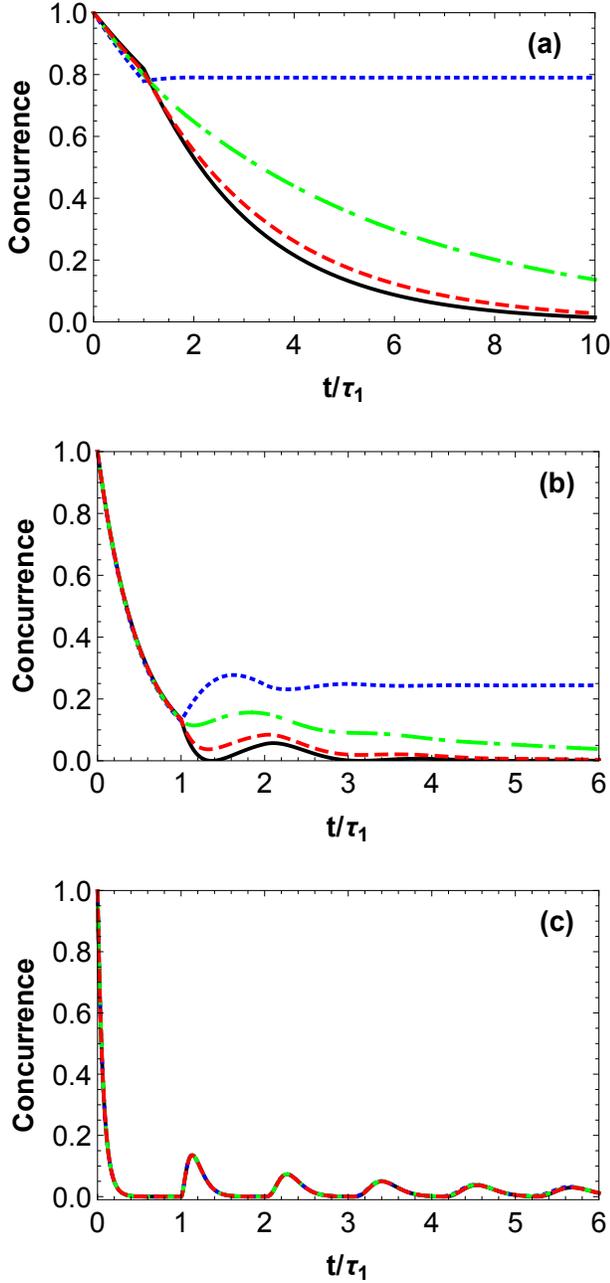}
\caption{(Color online) The concurrence between the TLSs as functions of the
dimensionless time $t/\protect\tau_{1}$ with initial condition $C_{s}(0)=1$
for different phase $\protect\varphi _{1}=2n\protect\pi$ (black solid
curve), $\protect\varphi _{1}=2n\protect\pi+\protect\pi$ (blue dotted
curve), $\protect\varphi _{1}=2n\protect\pi+\protect\pi/2$ (green dot-dashed
curve), $\protect\varphi _{1}=2n\protect\pi+\protect\pi/4$ (red dashed
curve) in (a)$n=2$, (b)$n=20$, (c)$n=150$. We have set the following
parameters: $a=2b$, $\protect\omega_{A}=(\Omega_{11}+\Omega_{31})/2$, $%
\protect\gamma _{1}\protect\lambda _{1}/v_{1}=0.05 $.}
\label{Fig2.eps}
\end{figure}
%%%%%%%%%%%%%%%%%%%%%%%%%%%%%%%%%%%%%%%%%%%%%%%%%%%%%%%%%%%%%%%%%%%%

To measure the amount of the entanglement, we use concurrence as the
quantifier~\cite{Wootters}. By taking a partial trace over the waveguide
degrees of freedom, the initial density matrix of the two TLSs is of an
X-form in the two-qubit standard basis $\{\left\vert 1\right\rangle
,\left\vert 2\right\rangle ,\left\vert 3\right\rangle ,\left\vert
4\right\rangle \}$. The concurrence for this type of state can be calculated
easily as
\end{subequations}
\begin{equation}  \label{2-A14}
C(t)=\max (0,2\left\vert B_{1}(t)B_{2}^{\ast }(t)\right\vert )
\end{equation}%
which can also expressed as the function of the amplitudes of the Dicke
states by the relation
\begin{subequations}
\label{2-A15}
\begin{eqnarray}
C_{s}(t) &=&\frac{B_{1}\left( t\right) +B_{2}\left( t\right) }{\sqrt{2}}, \\
C_{a}(t) &=&\frac{B_{1}\left( t\right) -B_{2}\left( t\right) }{\sqrt{2}}.
\end{eqnarray}
We expect that the position-dependent phase factors $e^{i\varphi _{j}}$ and
the delay times $\tau _{j}$ will lead to a modification of the entanglement
among the TLSs.

\subsection{Single transverse mode}

In the frequency band between $\Omega _{11}$ and $\Omega _{31}$, the
waveguide is said to be single-moded. The TLSs with the transition frequency
$\omega _{A}\in \left( \Omega _{11},\Omega _{31}\right) $ only emit photons
into the TM$_{11}$ ($j=1$) guided mode. In this case, the time behavior of
Dicke states reads
\end{subequations}
\begin{subequations}
\label{2-B1}
\begin{eqnarray}
C_{s}(t) &=&C_{s0}\sum_{n=0}^{\infty }\frac{\left( -\gamma _{1}e^{i\varphi
_{1}}\right) ^{n}}{n!}t_{n}^{n}e^{-\gamma _{1}t_{n}}, \\
C_{a}(t) &=&C_{a0}\sum_{n=0}^{\infty }\frac{\left( \gamma _{1}e^{i\varphi
_{1}}\right) ^{n}}{n!}t_{n}^{n}e^{-\gamma _{1}t_{n}}.
\end{eqnarray}%
where $t_{n}=t-n\tau _{1}$, and $C_{s0}$ and $C_{a0}$ are the initial
amplitude. The time axis is divided into intervals of length $\tau _{1}$. A
step character is presented in Eqs.(\ref{2-B1}). For $t\in \left[ 0,\tau _{1}%
\right] $, both amplitudes, $C_{s}$ and $C_{a}$, decay exponentially with
decay rate $\gamma _{1}$. The underlying physics is that one TLS requires at
least the time $\tau _{1}$ to recognize the other TLS. For $t\in \left[ \tau
_{1},2\tau _{1}\right] $, the absorption and reemission of light by each TLS
produce the interference, which results in a energy change between two TLSs.
%%%%%%%%%%%%%%%%%%%%%%%%%%%%%%%%%%%%%%%%%%%%%%%%%%%%%%%%%%%%
\begin{figure}[tbp]
\includegraphics[width=8cm]{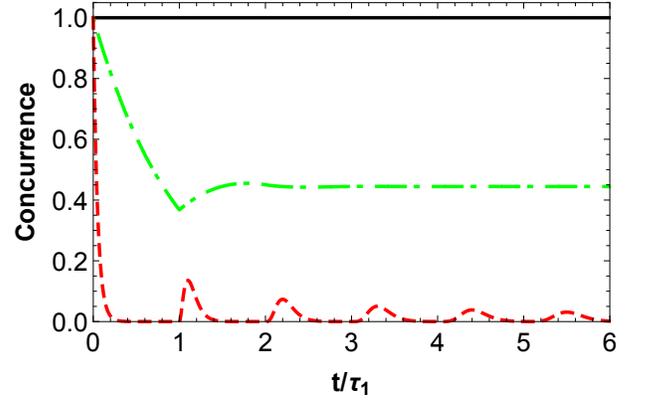}
\caption{(Color online) The concurrence between the TLSs as a function of
the dimensionless time $t/\protect\tau _{1}$ with initial condition $%
C_{a}(0)=1$ for distance $d=0$ (black solid line), $d=10\protect\lambda _{1}$
(green dot-dashed line), $d=200\protect\lambda _{1}$ (red dashed line).
Other parameters are the same as in Fig.~\ref{Fig2.eps}.}
\label{Fig3.eps}
\end{figure}
%%%%%%%%%%%%%%%%%%%%%%%%%%%%%%%%%%%%%%%%%%%%%%%%%%%%%%%%%%%%%%%%%%%%

From Eq.(\ref{2-B1}), one can observe that there is a $\pi $ phase
difference between the amplitudes $C_{s}(t)$ and $C_{a}(t)$. Hence, we
assume that the two TLSs are initially prepared in the symmetric state $%
C_{s0}=1$ to study the effect of the inter-TLS distance on the the dynamics
of entanglement between the TLSs. In this case, the concurrence takes the
maximum value between $0$ and $|C_{s}(t)|^{2}$. In Fig.~\ref{Fig2.eps}, we
have numerically plotted the concurrence as a function of time $t$ in units
of $\tau _{1}$ with $\gamma _{1}\lambda _{1}/\tau _{1}=0.05$, where the
wavelength $\lambda _{1}k_{10}=2\pi $. In the time interval $t\in \left[
0,\tau _{1}\right] $, two TLSs radiate spontaneously, so the concurrence
decays exponentially with time. As time goes on, the inter-TLS distance have
influenced the entanglement dynamics via phase $\varphi _{1}$ and delay time
$\tau _{1}$. When the two TLSs are close together, two TLSs act
collectively, the system dynamics is independent of the finite propagating
time of the light, which has been shown in Fig.~\ref{Fig2.eps}(a) with $%
\gamma _{1}\tau _{1}\ll 1$. There is stationary two-TLS entanglement when
the inter-TLS distance equals an odd integer number of $\lambda _{1}/2$,
(i.e., $\varphi _{1}=2n\pi +\pi $). And a small deviation of the special
position leads to the entanglement decaying asymptotically to zero. The
entanglement loses fast when the inter-TLS distance equals an integer number
of $\lambda _{1}$ corresponding to $\varphi _{1}=2n\pi $. For $\gamma
_{1}\tau _{1}\ll 1$, The dependence of the entanglement on phase in Fig.~\ref%
{Fig2.eps}(a) can be understood by letting $\tau _{1}\rightarrow 0$. In this
case, the amplitude of state $\left\vert s\right\rangle $ becomes
\end{subequations}
\begin{equation}
C_{s}(t)=C_{s0}\exp \left[ -t\gamma _{1}(1+\cos \varphi _{1})-it\gamma
_{1}\sin \varphi _{1}\right]  \label{2-B2}
\end{equation}%
It can be observed from Eq.~(\ref{2-B2}) that $|C_{s}(t)|$ exponentially
decays with time, it decays fast when $\varphi _{1}=2n\pi $ and keeps its
initial value when $\varphi _{1}=2n\pi +\pi $. Although one can explain the
relation of entanglement with phase by Eq.~(\ref{2-B2}), the probability of
finding the TLSs in the initial state is less than unity in Fig.~\ref%
{Fig2.eps}(a) when $\varphi _{1}=2n\pi +\pi $. Hence, the stationary two-TLS
entanglement indicates a superposition of the symmetry state in the absence
of photons and the ground TLS state in the presence of a photon. As the
inter-TLS separation increases a little bit to meet $\gamma _{1}\tau
_{1}\sim 1$ in Fig.~\ref{Fig2.eps}(b), both the decay time and the phase
play important roles due to the interference. The interference produced by
multiple reemissions and reabsorptions of photon results in an oscillatory
entanglement. Panel (c) of Fig.~\ref{Fig2.eps} illustrates the dynamics of
entanglement for a larger inter-TLS distance with $\gamma _{1}\tau _{1}\gg 1$%
. It can be observed that the phase does not make any sense. At early time,
each initially excited TLS emits light to the waveguide, the entanglement
begins to decrease abruptly from one. Then the concurrence keeps small and
approximates to zero until the time $t=\tau _{1}$, at which the excitation
get absorbed by each TLS. As soon as the emitted photon returns to the TLSs,
the entanglement is created. Then, the decrease of entanglement begins. The
entanglement of the TLSs exhibits peaks due to the iteration of the process
where a photon emitted by one of the atoms is reabsorbed by another atom,
but its periodic maxima are reduced in magnitude as $t$ increases because
the energy is carried away from TLSs by the forward-going waves emitted by
TLS 1 and the backward-going waves emitted by TLS 2. In this case, the
amplitudes are approximately described by
\begin{equation}
C_{s}(t)\propto \frac{\left( -\gamma _{1}e^{i\varphi _{1}}\right) ^{n}}{n!}
t_{n}^{n}e^{-\gamma _{1}t_{n}}  \label{2-BA}
\end{equation}%
in each time interval $\left[ n\tau _{1},\left( n+1\right) \tau _{1}\right]$%
.
\begin{figure*}[tbp]
\centering
\includegraphics[clip=true,width=18cm]{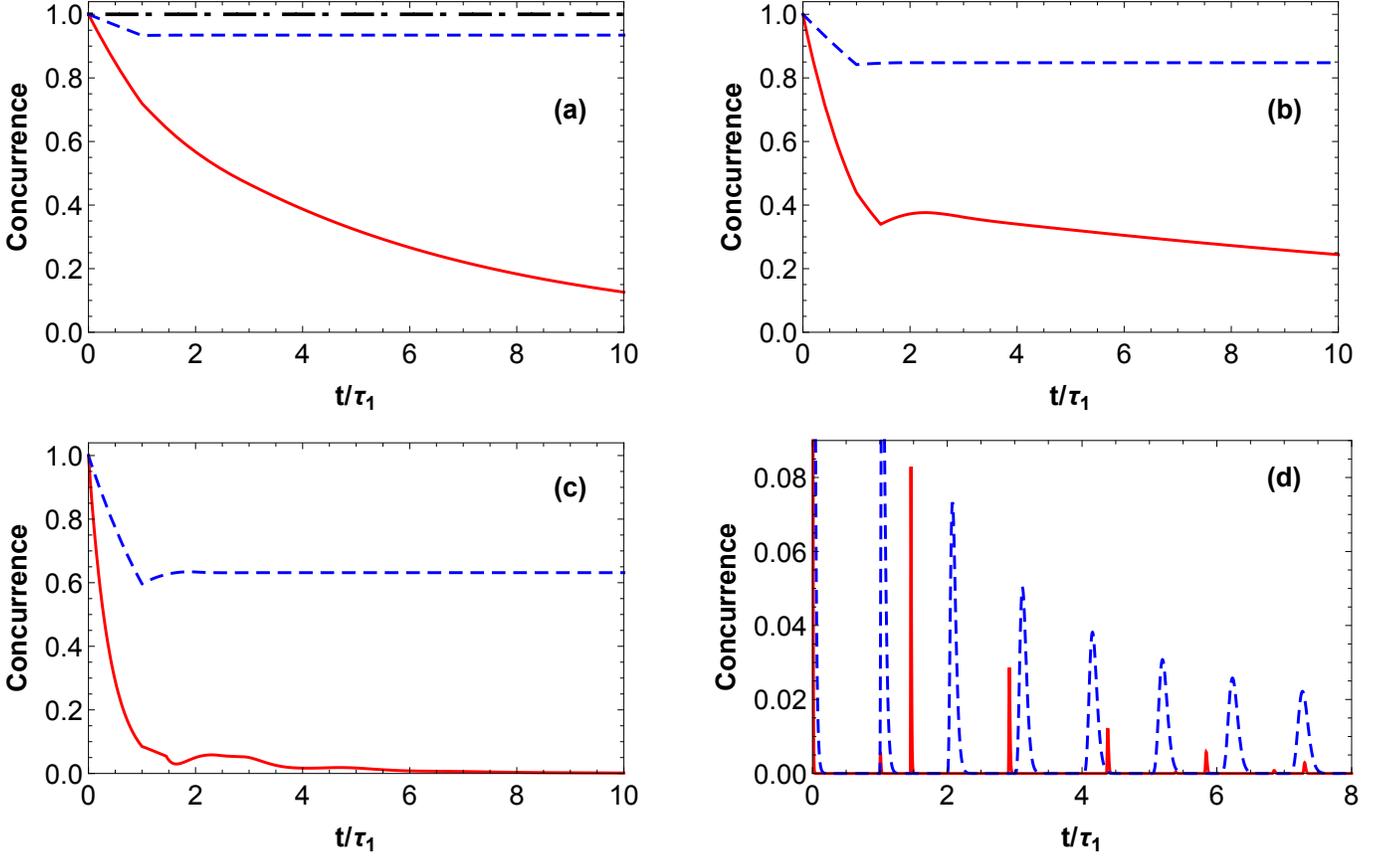}
\caption{(Color online) The time evolution of the concurrence between the
TLSs as a function of the dimensionless $t/\protect\tau _{1}$ with TLSs
initial in the antisymmetry state for phases $\protect\varphi _{1}=2n\protect%
\pi $ in (a) $n=4$, (b) $n=10$, (c) $n=30$, (d) $n=3000$. Here, the
concurrence for $d=0$ is given in the black dot-dashed lines. The
concurrence that only TM$_{11}$ mode is considered is presented by blue
dashed lines. The concurrence that both TM$_{11}$ and TM$_{31}$ modes are
considered is presented by the red solid lines. We have set the following
parameters: $a=2b$, $\protect\omega _{A}=(\Omega _{31}+\Omega _{51})/2$, $%
\protect\gamma _{1}\protect\lambda _{1}/v_{1}=0.0086$.}
\label{Fig4.eps}
\end{figure*}

For TLSs initially in the antisymmetry state, the concurrence exhibits the
similar behavior to the symmetry state with a $\pi$ phase difference.
However, when the inter-TLS spacing $d=0$, the antisymmetry state is a dark
state which does not interact with the electromagnetic field, it means the
the probability of finding the TLSs in the initial state is unity at any
time, so the concurrence is unchanged and remains its initial value one (see
the solid black line in Fig.~\ref{Fig3.eps}). We also plot the concurrence
as a function of the dimensionless time $t/\tau_{1}$ for phase $\varphi
_{1}=2n\pi $ with $n=10$ (the green dot-dashed line) and $n=200$ (red dashed
line) in Fig.~\ref{Fig3.eps}, where all the position-dependent phase factors
$e^{i\varphi_1}$ are equal. It can be seen that the cooperative effect
become lower and lower as the inter-TLS distance increases, so does the
maximum of concurrence.

\subsection{two transverse modes}

A TLS in its upper state radiates waves into the continua resonant with
itself. As the transition frequency of the TLSs increases, there are
additional guided modes taking part in the interaction with the TLSs. Due to
their different wave numbers, the inter-TLS distance introduces different
flight time $\tau _{j}$ of light between the TLSs as well as different
phases $\varphi _{j}$. From the definition of $\tau _{j}$ and $\varphi _{j}$%
, we know that $\tau _{j}<\tau _{j+1}$ and $\varphi _{j}<\varphi _{j+1}$ for
the given distance $d$. In this section, we assume that the transition
frequency of the TLSs is smaller than the cutoff frequency $\Omega _{51}$
and larger than the cutoff frequency $\Omega _{31}$, this means that the
emit photons will propagate in guided modes TM$_{11}$ and TM$_{51}$. The
delay-differential equation Eq.(~\ref{2-A13}) reduces to
\begin{subequations}
\label{2-B3}
\begin{eqnarray}
\left( \partial _{t}+\gamma \right) C_{s}(t) &=&-\alpha _{1}C_{s}\left(
t-\tau _{1}\right) \Theta \left( t-\tau _{1}\right)  \notag \\
&&-\alpha _{2}C_{s}\left( t-\tau _{2}\right) \Theta \left( t-\tau _{2}\right)
\\
\left( \partial _{t}+\gamma \right) C_{a}(t) &=&\alpha _{1}C_{a}\left(
t-\tau _{1}\right) \Theta \left( t-\tau _{1}\right)  \notag \\
&&+\alpha _{2}C_{a}\left( t-\tau _{2}\right) \Theta \left( t-\tau _{2}\right)
\end{eqnarray}%
where $\gamma =\gamma _{1}+\gamma _{2}$, and $\alpha _{j}=\gamma
_{j}e^{i\varphi _{j}}$ ($j=1,2$). We first discuss the case with $d=0$. It
is clear by inspection of Eq.~(\ref{2-B3}) that the initial entanglement
determined by the state $|s\rangle $ decreases more quickly in time, on the
contrary, the initial entanglement determined by the state $|a\rangle $ does
not change in time as shown by the black lines in Fig.~\ref{Fig4.eps}.

To analyze the influence of the number of the transverse modes on the
entanglement which interact with TLSs, we fix the phase $\varphi _{1}=2n\pi $
with $n=4$ in Fig.~\ref{Fig4.eps}(a), $n=10$ in Fig.~\ref{Fig4.eps}(b), $%
n=30 $ in Fig.~\ref{Fig4.eps}(c), $n=3000$ in Fig.~\ref{Fig4.eps}(d). We
plotted the time behavior of the concurrence between TLS by only considering
the TM$_{11}$ mode in Eq.(\ref{2-B3}), which is shown in blue dashed lines
in Fig.~\ref{Fig4.eps}. The red solid lines in Fig.~\ref{Fig4.eps} present
the time behavior of the concurrence by considering both TM$_{11}$ and TM$%
_{31}$ modes in Eq.~(\ref{2-B3}). It can be found that increasing the number
of the transverse modes which interact with TLSs leads to exponentially
decay with a rate $\gamma _{1}+\gamma _{2}$ up to time $t=\tau _{1}$. After
this, the phase $\varphi _{1}$ begins to have an effect until time $t=\tau
_{2}$. After time $\tau _{2}$, the dynamics can be dramatically affected by
the phases $\varphi _{j}$ and delay times $\tau _{j}$ induced by the
inter-TLS distance. The blue lines of panels (a) and (b) in Fig.~\ref%
{Fig4.eps} show that the concurrence remains constant in time after the time
$t=\tau _{1}$ in TM$_{11}$ mode, which indicate that the delay time $\tau
_{1}$ is negligibly small. As the phase factor $e^{i\varphi _{1}}$ is fixed,
the behavior of the red solid lines is completely determined by the phase $%
\varphi _{2}$ and delay time $\tau _{2}$ in TM$_{31}$ mode. We see from
panel (a) that the entanglement decays almost exponentially in time after
delay time $\tau _{2}$, which means that a part of the emitted energy from
one TLS is transferred directly to another TLS, so there is no delay in the
absorption of the energy by another TLS in both TM$_{11}$ and TM$_{31}$
modes. In this case, we can solve Eq.(\ref{2-B3}) by letting $\tau _{1},\tau
_{2}\rightarrow 0$
\end{subequations}
\begin{subequations}
\label{2-B4}
\begin{eqnarray}
C_{s}(t) &=&C_{s0}\exp \left[ -(\gamma _{1}+\alpha _{1})t-(\gamma
_{2}+\alpha _{2})t\right] , \\
C_{a}(t) &=&C_{a0}\exp \left[ -(\gamma _{1}-\alpha _{1})t-(\gamma
_{2}-\alpha _{2})t\right] .
\end{eqnarray}
The norm of amplitudes are completely determined by phases leading to
exponential decay of the entanglement. The red solid line in panel (b)
exhibits behavior different from that in panel (a), indicating that the $%
\varphi _{2}$ and delay time $\tau _{2}$ play an equal role. The part of
excitation emitted into TM$_{11}$ mode is immediately reabsorbed by the
other one, but the part of excitation emitted into TM$_{31}$ mode undergoes
delay, however, wave interference still produced at each exchanges of the
excitation between the TLSs in TM$_{31}$ mode. In this case, we can solve
Eq.(\ref{2-B3}) by letting $\tau _{1}\rightarrow 0$
\end{subequations}
\begin{subequations}
\label{2-B5}
\begin{eqnarray}
&&C_{s}=C_{s0}\sum_{n=0}^{\infty }\frac{\left( -\alpha _{2}\right) ^{n}}{n!}%
e^{-\left( \gamma +\alpha _{1}\right) \left( t-n\tau _{2}\right) }\left(
t-n\tau _{2}\right) ^{n} \\
&&C_{a}=C_{a0}\sum_{n=0}^{\infty }\frac{\alpha _{2}^{n}}{n!}e^{-\left(
\gamma -\alpha _{1}\right) \left( t-n\tau _{2}\right) }\left( t-n\tau
_{2}\right) ^{n}
\end{eqnarray}
Panel (c) shows that as $d$ increased, the delay time should be taken into
account in TM$_{11}$ mode besides the wave interference. In this case, we
can use Laplace transformation and geometric series expansion to solve Eq.(%
\ref{2-B3}), the solutions
\end{subequations}
\begin{subequations}
\label{2-B6}
\begin{eqnarray}
C_{s} &=&C_{s0}\sum_{n=0}^{\infty }\sum_{k=0}^{n}C_n^k\alpha _{1}^{k}\alpha
_{2}^{n-k}\frac{\left(\tau _{nk}-t\right) ^{n}}{n!}e^{-\gamma \left( t-\tau
_{nk}\right) }, \\
C_{a} &=&C_{a0}\sum_{n=0}^{\infty }\sum_{k=0}^{n}C_n^k\alpha _{1}^{k}\alpha
_{2}^{n-k}\frac{\left( t-\tau _{nk}\right) ^{n}}{n!}e^{-\gamma \left( t-\tau
_{nk}\right) },
\end{eqnarray}
are coherent sums over contributions starting at different instants of time $%
\tau _{nk}=k\tau _{1}+\left( n-k\right) \tau _{2}$, where $C_n^k=\frac{k!}{%
n!(n-k)!}$. Each term of the sum has a well-defined phase, and are damped by
an exponential function at rate $\gamma$. Interference is possible if the
amplitudes do not decay appreciably over the time $\tau_{nk}$.As $d$ is large enough so that $\min_{p,q}{\gamma|p\tau_2-q\tau_1|}\gg 1$ with non-negative integers $p$ and $q$ which are not zero at the same time,the phase factor plays no role as shown in panel (d). The wave packets of the emitted excitation is bouncing back and forth between two TLSs until its intensity is damped to zero. So there are collapses and revivals of the concurrence
until the amplitude of the revivals damped to zero.

%%%%%%%%%%%%%%%%%%%%%%%%%%%%%%%%%%%%%%%%%%%%%%%%%%%%%%%%%%%%

\section{\label{Sec:4} conclusion}

%%%%%%%%%%%%%%%%%%%%%%%%%%%%%%%%%%%%%%%%%%%%%%%%%%%%%%%%%%%%
We have studied the effects of the inter-TLS distance on the entanglement
properties of two identical TLSs located inside a rectangular hollow
metallic waveguide of transverse dimensions $a$ and $b$. When the energy
separation of the TLS is far away from the cutoff frequencies of the
transverse modes and there is single excitation in the system, the
Schrodinger equation for the wave function with single excitation initial in
the TLSs is reduced to the delay differential equations for the amplitudes
of two TLSs, where phase factors and delay times are induced by the finite
distance between the TLSs. The delay differential equations are solved
exactly for the TLSs interacting with either single transverse mode or
double transverse modes of the waveguide, which directly reveals the
retarded character of multiple reemissions and reabsorptions of photons
between the TLSs. For the inter-TLS distance $d=0$, there exists an
anti-symmetry state decoupled with the field modes, so the entanglement can
be generated if the TLSs are initial in a separate state, later trapped in
the anti-symmetry state. As the TLSs are close together such that the time
delay $\max \{\tau _{j}\}$ is much smaller than the TLS decay time $\gamma
^{-1}$, the excitation emitted by one TLS into the field is absorbed
immediately by the other. The dynamic of the entanglement are dramatically
affected by phases, leading to an enhanced and inhibited exponential decay
of the concurrence when only one transverse mode are considered, and an
exponential decay when more transverse modes are involved. As the inter-TLS
distance increases, both phases and delay times affect the concurrence.
There is a proper delay of reabsorption after reemission of photons but
interference is possible if the amplitudes of TLSs do not decay appreciably
over time $\tau_{nk}$. As $d$ is large enough so that $\min_{p,q}{\gamma|p\tau_2-q\tau_1|}\gg 1$ with non-negative integers $p$ and $q$ which are not zero at the same time,the phase factor plays no role. There appear collapses and revivals of the
entanglement of the TLSs. We note that our studies focus on the the
dependence of the concurrence on the inter-TLS distance but it is easy to
study the dependence of the concurrence on the initial state of the system
with the exact solution.

\begin{acknowledgments}
This work was supported by NSFC Grants No. 11434011, No. 11575058.
\end{acknowledgments}

\end{subequations}

\end{document}